\documentclass[twocolumn,showpacs,amsmath,amssymb,aps,preprintnumbers]{revtex4-1}

\usepackage{graphicx}
\usepackage[T1]{fontenc} 

\begin{document}

\preprint{Submitted to: ACTA PHYSICA POLONICA A}

\title{Charge orderings and phase separations
in the atomic limit \\ of the extended Hubbard model
 with intersite density-density interactions}

\author{Konrad Kapcia}%
    \email{kakonrad@amu.edu.pl}
\author{Waldemar K\l{}obus}
\author{Stanis\l{}aw Robaszkiewicz}%
\affiliation{ Electron States of Solids Division, Faculty of Physics, Adam Mickiewicz University, ul. Umultowska 85, 61-614 Pozna\'n, POLAND
}%

\date{April 14, 2010}

\begin{abstract}
A simple effective model of charge ordered insulators is studied. The tight binding Hamiltonian consists of the effective on-site interaction $U$ and the intersite density-density interactions $W_{ij}$ (both: nearest-neighbour and next-nearest-neighbour).
In the analysis of the phase diagrams we have adopted the variational approach, which treats the on-site interaction term exactly and the intersite interactions within the mean-field approximation.
The phase separated states have not been taken into account in previous analyses.
Our investigations of two cases of the on-site interaction: attraction (\mbox{$U/(-W_Q)=-10$}) and repulsion (\mbox{$U/(-W_Q)=1.1$}) show that, depending on the values of the next-nearest-neighbour attraction, the system can exhibit not only homogeneous phases: charge ordered (CO) and nonordered (NO), but also various phase separated states (CO--NO, CO--CO).

\end{abstract}


\maketitle

\section{Introduction}

Electron charge orderings phenomena are currently under intense investigations, because they are relevant to a broad range of important materials such as manganites, cuprates and organic conductors \cite{MRR1990,GL2003,SAC2001,J2004,SHF2004}. In this paper we will discuss an effective model of charge ordered insulators.

The Hamiltonian considered has the following form:
\begin{equation}
\label{row:1} \hat{H}  =  U\sum_i{\hat{n}_{i\uparrow}\hat{n}_{i\downarrow}} + \frac{W_{1}}{2}\sum_{\langle i,j\rangle_1}{\hat{n}_{i}\hat{n}_{j}}
+\frac{W_{2}}{2}\sum_{\langle i,j\rangle_2}{\hat{n}_{i}\hat{n}_{j}} - \mu\sum_{i}{\hat{n}_{i}},
\end{equation}
where $\hat{c}^{+}_{i\sigma}$ denotes the creation operator of an electron with spin $\sigma$ at the site~$i$, $\hat{n}_{i}=\sum_{\sigma}{\hat{n}_{i\sigma}}$, $\hat{n}_{i\sigma}=\hat{c}^{+}_{i\sigma}\hat{c}_{i\sigma}$,
$U$~is the on-site density interaction,
$W_{1}$ and $W_{2}$ are the intersite density-density interactions between nearest neighbours
and next-nearest neighbours, respectively. $\mu$ is the chemical potential, depending on the concentration of electrons:
\begin{equation}\label{row:2}
n = \frac{1}{N}\sum_{i}{\left\langle \hat{n}_{i} \right\rangle},
\end{equation}
with \mbox{$0\leq n \leq2$} and $N$ is the total number of lattice sites.

The interactions $U$ and $W_{ij}$ will be treated as the effective ones and will be assumed to include all the possible contributions and renormalizations like those coming from the strong electron-phonon coupling or from the coupling between electrons and other electronic subsystems in solid or chemical complexes. In such a general case arbitrary values and signs of $U$ and $W_{ij}$ are important to consider.

We have performed extensive study of the phase diagrams of the model (\ref{row:1}) for arbitrary $n$~\cite{K2009,KKR2009}.
In the analysis we have adopted a variational approach (VA) which treats the on-site interaction $U$ exactly and the intersite interactions ($W_{ij}$) within the mean-field approximation (MFA).
Within such an approach the phase diagrams of (\ref{row:1}) have been investigated
till now for the special case \mbox{$W_2=0$} only~\cite{MRC1984,R1979}.

In the following we will restrict ourselves to the case of repulsive \mbox{$W_1>0$} and attractive \mbox{$W_2<0$}. We consider only two-sublattice orderings on the lattice consisting of two interpenetrating sublattices such as for example sc or bcc lattices.

    \begin{figure*}
        \centering
        \includegraphics[width=0.9\textwidth]{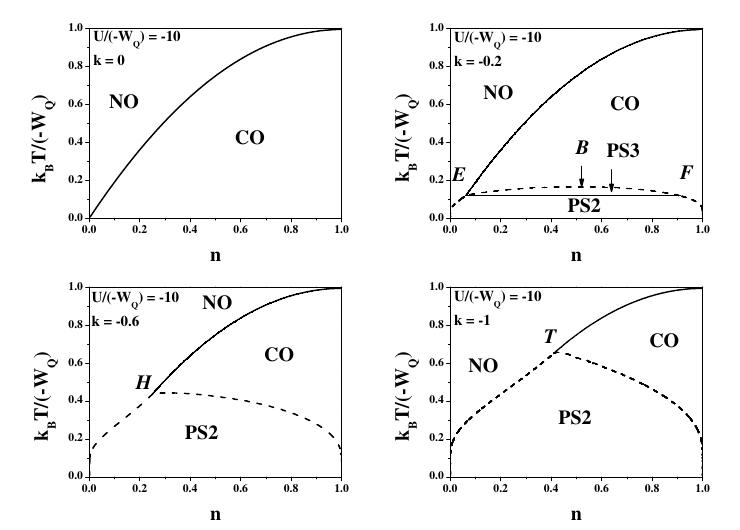}
        \caption{Phase diagrams $k_BT/(-W_Q)$~vs.~$n$ for \mbox{$U/(-W_Q)=-10$}, \mbox{$W_1>0$} and \mbox{$k = z_2W_2/z_1W_1 = 0,\ -0.2,\ -0.6,\ -1$} (as labeled). Solid and dashed lines indicate second order and ``third order'' boundaries, respectively.}
        \label{rys:Um10}
    \end{figure*}

Within the VA the intersite interactions are decoupled within the MFA,
what let us find a free energy per site $f(n)$.
The condition (\ref{row:2}) for the electron concentration and a~minimization of $f(n)$ with
respect to the charge-order parameter lead to a set of two self-consistent equations (for homogeneous phases), which are solved numerically.
The charge-order parameter is defined as \mbox{$n_Q=(1/2)(n_A-n_B)$}, where $n_{\alpha}=\frac{2}{N}\sum_{i\in\alpha}{\left\langle\hat{n}_i\right\rangle}$ is the average electron density in a sublattice \mbox{$\alpha=A,B$}. If $n_Q$ is non-zero the charge-ordered phase (CO) is a solution, otherwise the non-ordered phase (NO) occurs.

Phase separation (PS) is a state in which two domains with different electron concentrations exist in the system
(coexistence of two homogeneous phases). The free energies of the PS states are calculated from the expression:
\begin{equation}
f_{PS}(n_{+},n_{-}) = m f_{+}(n_{+}) + (1-m) f_{-}(n_{-}),
\end{equation}
where $f_{\pm}(n_{\pm})$ are values of a free energy at $n_{\pm}$ corresponding to the
lowest energy homogeneous solutions and
\mbox{$m  = \frac{n - n_-}{n_+ - n_-}$}
is a fraction of the system with a charge density $n_+$.
We find numerically the minimum of $f_{PS}$ with respect to $n_+$ and $n_-$.

In the model considered only the following PS states can occur: PS2 is a coexistence of CO and NO phases and PS3 is a coexistence of two CO phases with different concentrations (and charge-order parameters).

In the paper we have used the following convention. A~second order transition is a~transition between homogeneous phases with a~continuous change of the order parameter at the transition temperature. A~transition between homogeneous phase and PS state is symbolically named as a~``third order'' transition. During this transition a~size of one domain in the PS state decreases continuously to zero at the transition temperature.  We also distinguished second order transition between two PS states, at which a continuous change of the order parameter in both domains takes place.

Second order transitions are denoted by solid lines on phase diagrams and dashed lines correspond to the ``third order'' transitions. We introduce also the following denotation: \mbox{$W_Q=-z_1W_1+z_2W_2$}, where $z_1$ and $z_2$ are numbers of nearest and next-nearest neighbours, respectively.

Obtained phase diagrams are symmetric with respect to half-filling (\mbox{$n=1$}) because of the particle-hole symmetry of the Hamiltonian (\ref{row:1}), so the diagrams will be presented only in the range \mbox{$0\leq n\leq 1$}.

\section{Results and discussion}

    \begin{figure*}
        \centering
        \includegraphics[width=0.9\textwidth]{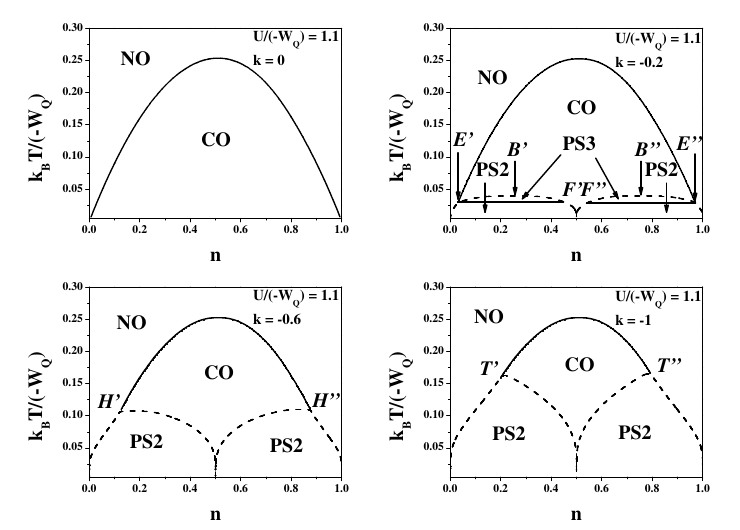}
        \caption{Phase diagrams $k_BT/(-W_Q)$~vs.~$n$ for \mbox{$U/(-W_Q)=1.1$}, \mbox{$W_1>0$} and \mbox{$k = 0,\ -0.2,\ -0.6,\ -1$} (as labeled). Solid and dashed lines indicate second order and ``third order'' boundaries, respectively.}
        \label{rys:U11}
    \end{figure*}

    Examples of the $k_BT$~vs.~$n$ phase diagrams evaluated for \mbox{$U/(-W_Q)=-10$}, \mbox{$W_1>0$} and various ratios of \mbox{$k=z_2W_2/z_1W_1\leq 0$} are shown in Fig.~\ref{rys:Um10}.
    If \mbox{$0\leq|k|<1$} the CO and NO (homogeneous) states are separated by the second order transition line.

    When \mbox{$-0.6<k<0$} a ``third order'' transition takes place at low temperatures, leading first to PS into two coexisting CO phases (PS3), while at still lower temperatures CO and NO phases coexist (PS2). The critical point (denoted as $B$) for this phase separation is located inside the CO phase. The \mbox{$E$-$F$} solid line is associated with continuous transition between two different PS states (\mbox{PS2--PS3}, the second order \mbox{CO--NO}  transition occurs in the domain with lower concentration).

    For \mbox{$k<-0.6$} the transition between PS states does not occur, the area of PS3 stability vanishes and the critical point for the phase separation (denoted as $T$) lies on the second order line \mbox{CO--NO}. As \mbox{$k \rightarrow -\infty$} \mbox{$T$-point} occurs at \mbox{$n=1$} and the homogeneous CO phase does not exist beyond half-filling.

    When \mbox{$k=-0.6$} the lower branch of the ``third order'' curve approaches the critical point ($H$) parabolically. \mbox{$H$-point} is a higher order critical point and at this point the lines consisting of $E$, $F$ and $T$ points connect together.

    For \mbox{$U/(-W_Q)=1.1$} sequences of transitions are similar to the previous case (for corresponding values of $k$), but now the phase diagrams are (almost) symmetric with respect to quarter-filling (\mbox{$n=0.5$}). $B'$, $H'$, $T'$, $E'$ and $F'$ points (as well as $B''$, $H''$, $T''$, $E''$ and $F''$ points) appear, which correspond to $B$, $H$, $T$, $E$ and $F$ points, respectively. The obtained phase diagrams are shown in Fig.~\ref{rys:U11}.

\section{Conclusions}

In this paper we studied atomic limit of the extended Hubbard model with intersite repulsion \mbox{$W_1>0$} and next-nearest neighbour attraction \mbox{$W_2<0$}. We considered two qualitatively different regimes of the on-site interaction: strong attraction \mbox{$U/(-W_Q)=-10$} and (relatively) strong repulsion \mbox{$U/(-W_Q)=1.1$}. Our analyses show that for attractive $W_2$ and \mbox{$n\neq 1$} the states with phase separation have the lowest free energy at sufficiently low temperatures \mbox{$T\geq0$}, whereas for \mbox{$W_2=0$} only homogeneous phases exist on the phase diagrams which have the form presented in Ref.~\cite{MRC1984}.

The areas of PS states stability expand with increasing of the next-nearest neighbour attraction strength. Moreover, the continuous transition between two different PS states occurs for \mbox{$0<|k|<0.6$}.
One should notice that a change of the strength of the next-nearest attraction can modify a type of the critical point for separation (which can be $B$, $T$ or $H$-point).

\end{document}